\documentclass[twocolumn,pre]{revtex4}
\usepackage{epsfig}
\usepackage{graphicx}
\usepackage{dcolumn}
\usepackage{bm}

\def\be{\begin{equation}}
\def\ee{\end{equation}}

\def\lan{\langle}
\def\ran{\rangle}

\begin{document}


\title{Thermodynamics and Statistical Mechanics of frozen systems in inherent states}
\author{Annalisa Fierro}
\author{Mario Nicodemi}
\altaffiliation[Also at ]{Department of Mathematics, Imperial College, London,
SW7 2BZ, U.K.}
\author{Antonio Coniglio}
\affiliation{Dipartimento di Fisica, Universit\`{a} di Napoli ``Federico II'',
INFM, Unit\`{a} di Napoli, Complesso Universitario Monte Sant'Angelo, Via
Cinthia, I-80126, Napoli, Italy}
\date{\today}
\begin{abstract}
We discuss a Statistical Mechanics approach in the manner of Edwards to the 
``inherent states'' (defined as the stable configurations in the potential 
energy landscape) of glassy systems and granular materials.
We show that at stationarity the inherent states are distributed
according a generalized Gibbs measure obtained assuming the validity of the
principle of maximum entropy, under suitable constraints.
In particular we consider three lattice models (a diluted Spin Glass, a
monodisperse hard-sphere system under gravity and a hard-sphere binary mixture
under gravity)
undergoing a schematic ``tap dynamics'', showing via Monte Carlo
calculations that the time average of macroscopic quantities over the tap
dynamics and over such a generalized distribution coincide.
We also discuss about the
general validity of this approach to non thermal systems.
\end{abstract}
\maketitle

\section{Introduction}

There are many complex
systems where thermal fluctuations are small enough that the temperature
of the external bath, $T_{bath}$, can be considered zero. 
Examples are supercooled
liquids quenched at zero temperature in metastable states (blocked
configurations), called ``inherent structures'', which correspond to the
local minima of the potential energy in the particles configuration space 
\cite{Stillinger,parisi,sciortino,tartaglia}. Granular materials 
\cite{JNBHM} at rest are another important example of system frozen 
\cite{nota_frozen} in
mechanically stable microstates (blocked configurations), which by analogy
with the glass terminology can also be called inherent states. 

The issue we consider here, which recently raised considerable 
interest, is to investigate the possibility to describe these systems 
by using concepts from Statistical Mechanics, as Edwards \cite{Edwards} 
suggested for granular media more than ten years ago. 
His assumption was that, by gently shaking the system under the
constrain  of fixed volume $V$,  the distribution over the mechanically
stable (blocked) states would be uniform. This leads to the definition
of a configurational entropy, $S = \ln \Omega$, where $\Omega$ is the number
of mechanically stable states corresponding to the fixed volume $V$ and
energy $E$, and to the concept of compactivity, $X^{-1} =
\frac{\partial \ln \Omega}{\partial V}$. In a similar way one can also
define a configurational temperature, $T_{conf}^{-1} \equiv \beta_{conf}=
\frac{\partial \ln \Omega}{\partial E}$.

Also in glasses, following for example the inherent structure approach 
\cite{Stillinger,parisi,sciortino,tartaglia}, one can
define  a configurational entropy associated to the number of
inherent structures corresponding to a fixed energy, $E$, and consequently the
configurational temperature. When the system is frozen at zero temperature
in one of his inherent states it does not evolve anymore. However, one can
explore the inherent structures space essentially in two ways. One way
is by quenching  the system over and over from an equilibrium temperature,
T, to zero temperature \cite{Stillinger,sciortino,tartaglia}.
Using this procedure, Sciortino {\em et al.}  \cite{sciortino} 
found that in a supercooled
glass forming liquid, studied by molecular dynamics simulations, the
configurational temperature
numerically coincides with the equilibrium temperature $T$,  
provided that $T$ is low enough.  

Another way to visit the inherent structures is by letting the system
aging in contact with an almost zero bath temperature, $T_{bath}$. 
During the aging
process an effective temperature, $T_{dyn}$, can be defined via the
off-equilibrium extension of the Fluctuation-Dissipation ratio \cite{peliti}.
It happens that in mean field models \cite{franzvirasoro} this effective
temperature coincides with the above configurational temperature. The
possibility to introduce an effective temperature for granular media via
the extension of the Fluctuation-Dissipation relation, was suggested in
Ref. \cite{n1}.

The connection between Edwards approach for granular media and the results
in glass theory has been pointed out in \cite{n1,barrat,Makse} 
and \cite{NC,NCF1}. 
In particular in Ref. \cite{barrat} it was shown that, for a class of finite
dimensional systems, in the limit $T_{bath}\rightarrow 0$, 
$T_{dyn}$ coincides in fact with the
configurational temperature, predicted by the Edwards' hypothesis.


In Ref. \cite{NC,NCF1} the inherent states are visited in an other way 
by using a tap 
dynamics (i.e., a procedure similar to that used in the compaction of
real granular materials), where each tap consists in raising the bath
temperature to a value $T_{\Gamma}$ and, after a lapse of time $\tau_0$,
quenching it back to zero. By cyclically repeating the process  the system
explores the space of the inherent states
\cite{NC,NCF1,NFC,brey,Dean,berg,mehta,NCH}. 
Once the stationary state is reached
one can define a temperature, $T_{fd}$, via the equilibrium 
Fluctuation-Dissipation relation.
One can then see that, 
if Edwards' assumption applies, $T_{fd}$ coincides with the configurational
temperature. This has been verified in fact for different finite
dimensional models \cite{NC,NCF1,NFC}. 
It was also shown numerically that for low 
enough $T_{\Gamma}$ one has that $T_{fd} =T_{conf}\simeq T_{\Gamma} $,
confirming on lattice models for granular media the result of 
Ref. \cite{sciortino}. 
In fact when the duration of each single tap is infinite 
($\tau_0\rightarrow\infty$), the tap coincides with the
way to explore the inherent states implemented in molecular dynamics
simulations for Lennard Jones mixtures \cite{Stillinger,sciortino}. However
the method used in \cite{sciortino} only allows the calculation of $T_{conf}$ 
when the configurational temperatures is low, i.e., 
where all the different temperatures almost coincide. 
Many other studies confirming Edwards' approach have also
been presented \cite{brey,Dean,berg,cugliandolo}.

In this paper we give a comprehensive view of the results obtained in
\cite{NC,NCF1,NFC} by considering other models and giving more details.
In particular we study here three schematic lattice models for glassy
systems and granular media, i.e., a diluted Spin Glass, a monodisperse
hard-sphere system under gravity and a hard-sphere binary mixture under
gravity. In particular, in the diluted Spin Glass and in the monodisperse
hard-sphere system under gravity, the asymptotic states reached by the
system are found to be described only by the configurational temperature. 
Whereas in the hard-sphere binary mixture under gravity the asymptotic
states are found to be described by two
thermodynamic parameters \cite{nota_2t}, coinciding with the two
configurational temperatures which characterize the distribution among the
inherent states when the principle of maximum entropy is satisfied under
the constraint that the energies of the two species are independently
fixed. In Ref. \cite{NFC} a description of the segregation observed in the
binary system in term of these two temperatures is also given.

In Sect.s \ref{model} and \ref{tapping}, the frustrated lattice gas model
and the results of its study with the tap dynamics are
respectively presented. In Sect. \ref{outequ}, the same results are
obtained at higher density where the system at small temperature reaches a
quasi-stationary state in which one time quantities decay as the logarithm
of time. In Sect. \ref{EDW} Edwards' hypothesis is formulated using the
principle of maximum entropy. The  results obtained in the monodisperse
hard-sphere system under gravity are shown in Sect. \ref{hardsphere}.
In Sect. \ref{twotemp} the Statistical Mechanics approach is
extended to the hard-sphere binary mixture under gravity, where two
thermodynamic parameters are necessary to describe the asymptotic states
reached by the system.
Finally, in the Conclusions we draw a picture of the Statistical Mechanics 
approach to systems found in inherent states, as it emerges from our extensive
investigation. 

\section{The frustrated lattice gas model}
\label{FLG}
\subsection{The model}
\label{model}
Recently a lattice model has been introduced to describe glass formers
\cite{dfnc,varenna} and, in presence of gravity, granular materials
\cite{NCH,our_rev,jef}. The Hamiltonian of the model is:
\be
-H = J\sum_{\lan ij \ran}
(\epsilon_{ij}S_i S_j - 1)n_in_j +\mu \sum_i n_i,
\label{flg}
\ee
where the sum $\sum_{\lan ij \ran}$ is over nearest neighbor sites,
$S_i={\pm} 1$ are Ising spins, $n_i=0,1$ are occupation variables, $\mu$ is the
particle chemical potential and $\epsilon_{ij}$ are quenched and random
variables, equal to ${\pm} 1$ with equal probability.
This model reproduces the Ising spin glass in the limit
$\mu\rightarrow\infty$ (i.e., when all sites are occupied, $n_i\equiv 1$).

In the other limit, $J\rightarrow \infty$,
the model describes a frustrated lattice gas with properties
recalling those of a ``frustrated'' liquid.
In fact the first term of Hamiltonian (\ref{flg}) implies that two
nearest neighbor sites can be freely occupied only if their spin variables
satisfy the interaction, that is if $\epsilon_{ij}S_iS_j=1$, otherwise
they feel a strong repulsion.  To make the connection with a liquid, we note
that the internal degree of freedom, $S_i$, may represent for example the
internal orientation of a non spherical particle.
Two particles can be nearest neighbors only if the relative orientation
is appropriate, otherwise they have to move apart.
Since in a frustrated loop the spins cannot satisfy all interactions, in this
model particle configurations in which a frustrated loop is fully occupied are
not allowed.  The frustrated loops in the model are the same of the spin glass
model and correspond in the liquid to those loops which, due to geometrical
hindrance, cannot be fully occupied by the particles.
In $3D$ \cite{varenna,antonio}, the model has a maximum density
$\rho_{\text{max}}\simeq 0.68$, and a transition at $\rho_c\simeq 0.62$
where the non-linear spin susceptibility diverges.

In the present paper, the $3D$ cubic 
frustrated lattice gas model with $J$ finite
is considered.
The value of particle density, $\rho=\sum_i n_i/L^3$ ($L$ is the lattice
linear size), is fixed, and a Monte Carlo tap
dynamics, which allows the system
to explore its inherent states, is applied.
During the dynamics,
the system cyclically evolves for a time $\tau_0$
(the tap duration \cite{nota1}) at a finite value of the bath temperature,
$T_{\Gamma}$ (the tap amplitude), and afterwards 
it is suddenly frozen at zero temperature in one of its
inherent states (at zero temperature the system does not evolve anymore if the energy cannot be decreased by one single particle movement).
After each tap, when the system is at rest, we record the
quantities of interest. The time, $t$, considered is therefore discrete and
coincides with the number of taps.

\subsection{The results obtained under the tap dynamics}
\label{tapping}
\begin{figure}
\mbox{\epsfysize=7cm\epsfbox{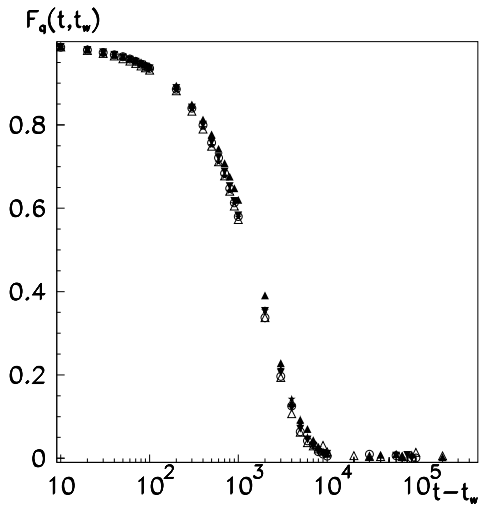}}
\caption{\label{dinamica} The self-scattering two time
function, $F_q(t,t_w)=\sum_i e^{\vec q{\cdot}[\vec r_i(t)-\vec r_i(t_w)]}/\rho
L^3$, with $q=\pi/4$, as  function of $t-t_w$
(for $t_w=10^4$, $2{\cdot} 10^4$, $5{\cdot} 10^4$,
$8{\cdot} 10^4$, $10^5$)
in the frustrated lattice gas model for density
$\rho=0.65$, during the tap dynamics, with tap amplitude
$T_{\Gamma}=0.3~J$ and tap duration $\tau_0=1~MCS$.
The function $F_q(t,t_w)$ only depends on $t-t_w$, showing that the system 
has reached stationarity. 
}
\end{figure}

\begin{figure}
\mbox{\epsfysize=7cm\epsfbox{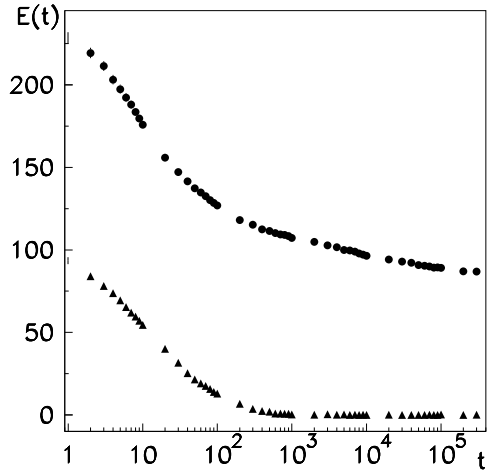}}
\caption{\label{energia}The energy, $E(t)$,
of the inherent states as  function of the tap number $t$,
in the frustrated lattice gas model during the
tap dynamics with tap amplitude $T_{\Gamma}=0.3~J$ and tap duration
$\tau_0=1~MCS$.
The lower curve, corresponding to a density of particles $\rho=0.65$, 
exponentially saturates to its asymptotic value, whereas the upper curve, 
corresponding to $\rho=0.75$, shows a logarithmic relaxation at long times. 
}
\end{figure}

We first consider the case $\rho=0.65$ \cite{nota2}. At this value of the
density the system under the tap dynamics reaches a stationary state for
each value of $T_{\Gamma}$ (and $\tau_0$) considered.
In Fig.s \ref{dinamica} and \ref{energia} \cite{nota3}, 
the self-scattering two time
function, $F_q(t,t_w)=\sum_i e^{\vec q {\cdot} [\vec r_i(t)-\vec
r_i(t_w)]}/\rho L^3$, and the energy, $E(t)$, of the inherent states,
obtained for $T_\Gamma=0.3~J$ and $\tau_0=1~MCS$, are shown. 
The curves, $F_q(t,t_w)$, for different $t_w$, collapse onto a single master
function, when they are plotted as function of $t-t_w$, and the energy,
$E(t)$, reaches its time independent asymptotic value, showing that the system
has reached a stationary state (our data are averaged up to $32$ noise
realizations; $L=8$ and $q=\pi/4$).


\begin{figure}
\mbox{\epsfysize=7cm\epsfbox{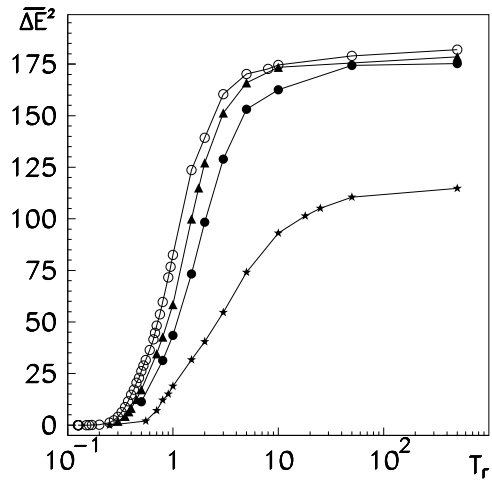}}
\caption{\label{fluctuation}
The time average of energy fluctuations, $\overline {\Delta E^2}$,
recorded in the stationary regime, as function of the tap amplitude,
$T_\Gamma$ (in units of $J$), in the frustrated lattice gas model for 
$\rho=0.65$.
The four different curves correspond to different values of the tap duration,
$\tau_0=~1,~5,~10,~\infty~MCS$ (from bottom to top).
This shows again that $T_{\Gamma}$ is not a right thermodynamic parameter. 
}
\end{figure}

During the tap dynamics, in the stationary state, the
time average of the energy, ${\overline E}$, and its fluctuations, ${\overline
{\Delta E^2}}$, are calculated.
\begin{figure}
\mbox{\epsfysize=7cm\epsfbox{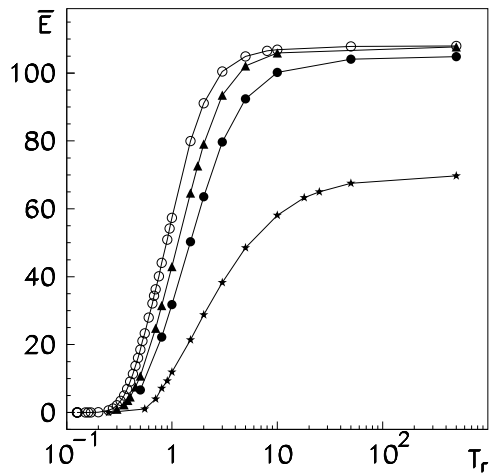}}
\caption{\label{energy} The time average of the energy, $\overline {E}$,
recorded in the stationary regime, as function of the tap amplitude,
$T_\Gamma$ (in units of $J$), in the frustrated lattice gas model for
$\rho=0.65$.
The four different curves correspond to
different values of the tap duration,
$\tau_0=~1,~5,~10,~\infty~MCS$ (from bottom to top).
This shows that $T_{\Gamma}$ is not a
right thermodynamic parameter, since sequences of taps with
different $\tau_0$ give different values for the system observables. 
}
\end{figure}
In Fig.s \ref{energy} and \ref{fluctuation},
${\overline E}$ and ${\overline{\Delta E^2}}$ are shown as
function the tap amplitude, $T_\Gamma$, (for several values of the tap
duration, $\tau_0$). Apparently, $T_{\Gamma}$ is not the right
thermodynamic parameter, since sequences of taps, with same $T_{\Gamma}$ and
different $\tau_0$, give different values of ${\overline E}$ and
${\overline {\Delta E^2}}$. On the other hand,
if the stationary states are indeed characterized by
a {\em single} thermodynamic parameter, $\beta_{fd}$,
the curves corresponding to different tap sequences (i.e. different
$T_{\Gamma}$ and $\tau_0$) should collapse onto a single master function, when
${\overline {\Delta E^2}}$ is parametrically plotted as function of
${\overline E}$. This data collapse is in fact found and shown in
Fig. \ref{universal}. This is a prediction which
could be easily checked in real granular materials (where one could consider
the density which is easier to measure than the energy).

\begin{figure}
\mbox{\epsfysize=7cm\epsfbox{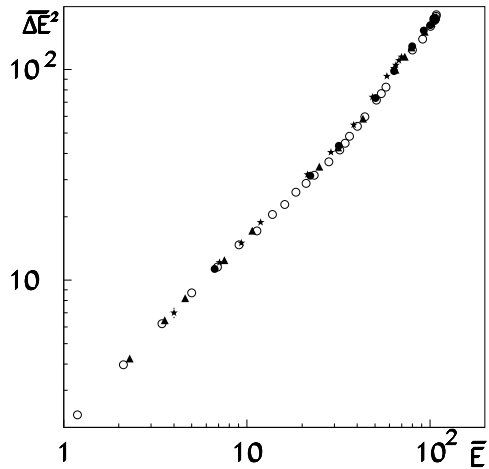}}
\caption{\label{universal}
The time averages of energy fluctuations, $\overline {\Delta E^2}$, 
when plotted as function of the time average of energy, $\overline {E}$,
collapse onto a single master function, for all the different
values of tap amplitude and duration, $T_\Gamma$ and $\tau_0$, plotted 
in Fig.\ref{energy}. 
This shows that the system stationary states are indeed characterized by
a {\em single} thermodynamic parameter, since 
the curves corresponding to different tap sequences (i.e. different
$T_{\Gamma}$ and $\tau_0$) collapse on a ``universal'' function, when
${\overline {\Delta E^2}}$ is parametrically plotted as function of
${\overline E}$. 
}
\end{figure}

The thermodynamic parameter, $\beta_{fd}$, is defined apart from an
integration constant, $\beta_0$, through the usual equilibrium 
Fluctuation-Dissipation relation:
\begin{equation}
-\frac{\partial {\overline E} }{\partial \beta_{fd}}  =
{\overline {\Delta E^2}}.
\label{Sb}
\end{equation}
By integrating eq. (\ref{Sb}), $\beta_{fd}-\beta_0$ can be expressed as
function of ${\overline E}$ or (for a fixed value of $\tau_0$) as function
of $\beta_\Gamma$: $ \beta_{fd}-\beta_0\equiv g(\beta_\Gamma)$. 
The constant $\beta_0$ is determined as explained in detail 
in the Appendix.

\subsection{The Edwards' averages}
\label{EDW}
In Sect. \ref{tapping} we have found that the fluctuations of the energy
in the stationary state depend only on the energy, $\overline{E}$,
and not on the past
history. If all macroscopic quantities depend only on the energy,
$\overline{E}$,
or on its conjugate thermodynamic parameter, $\beta_{fd}$,
the stationary state
can be  genuinely considered a ``thermodynamic state''. In this case
one can attempt to construct an equilibrium statistical mechanics, 
as originally suggested by Edwards \cite{Edwards}.

More precisely one can try to find from basic general principles 
what is the probability distribution, $P_r$,
of finding, in the stationary regime, 
the system in the inherent state $r$ of energy $E_r$ (see
\cite{NC}).  We assume that the distribution is given by the principle of
maximum entropy, $S=-\sum_r P_r
\ln P_r$, under the condition that the average energy is fixed: $E =
\sum_r P_r E_r $. Thus, we have to maximize the following
functional: $ I[P_r] =-\sum_r P_r
\ln P_r -\beta_{conf} (E - \sum_r P_r E_r) $.  Here $\beta_{conf}$ is a
Lagrange multiplier determined by the constraint on the energy and takes
the name of ``inverse configurational temperature''. This procedure leads
to the Gibbs result:
\begin{eqnarray}
P_r=\frac{e^{-\beta_{conf} E_r}}{Z}
\label{pr}
\end{eqnarray}
where $Z=\sum_r e^{-\beta_{conf} E_r}$. Using standard Statistical Mechanics
it is easy to show that, in the
thermodynamic limit, the entropy $S$  and $\beta_{conf}$ are also given by:
\begin{eqnarray}
S = \ln \Omega (E),\quad \beta_{conf}= \frac{\partial \ln \Omega}{\partial E}
\label{omega}
\end{eqnarray}
where $\Omega (E)$ is the number of inherent states corresponding to
energy $E$.

\begin{figure}
\mbox{\epsfysize=7cm\epsfbox{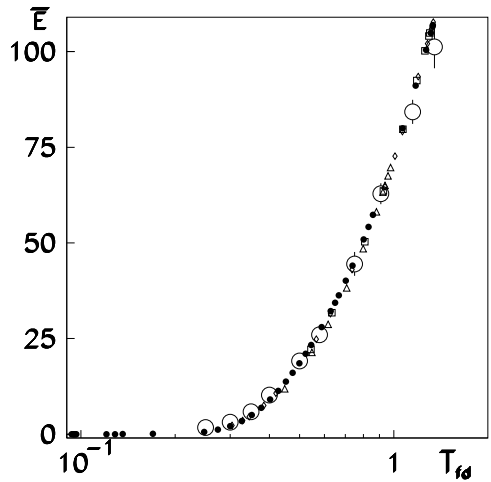}}
\caption{\label{edwards}
The time average, ${\overline E}$, calculated in the stationary regime of 
the tap dynamics and the ensemble average over the Edwards' distribution 
eq.(\ref{pr}),
$\lan E \ran$ (the black empty circles), are plotted respectively
as a function of $T_{fd}$  and $T_{conf}$
(in units of $J$), in the frustrated lattice
gas model, at $\rho=0.65$.
The two independently calculated sets of points show a very good agreement, 
outlining the success of Edwards' approach to describe the system 
macroscopic properties. 
}
\end{figure}

If the distribution in the stationary state coincides with eq.(\ref{pr})
the time average of the energy, ${\overline  E}(\beta_{fd})$,
recorded during the taps sequences, must coincide with the ensemble average,
$\lan E\ran(\beta_{conf})$, over the distribution eq.(\ref{pr}). In order
to check that we have independently calculated the average $\lan E\ran$, as
function of $\beta_{conf}$. We have simulated the model
eq.(\ref{flg}) imposing that the only accessible states are the
inherent states, as done in Ref. \cite{barrat}. 
The only difference is that in the present paper the Edwards' averages are
done in the canonical ensemble, whereas in Ref. \cite{barrat} are done
in the microcanonical ensemble.   
In particular we have constructed a new Hamiltonian, ${\cal
H'}(\{S_i, n_i\}) = {\cal H}(\{S_i, n_i\}) + \delta(\{S_i, n_i\})$,
by adding a term to eq.(\ref{flg}), $\delta(\{S_i, n_i\})$, which is zero,
if the configuration is an inherent state, and infinite, otherwise. The
canonical distribution for this Hamiltonian gives a weight,
$e^{{-\beta_{conf}}{\cal H'}}$, which coincides with the weight in the
distribution of eq.(\ref{pr}), for each accessible configuration.
Using the standard Monte Carlo simulations, we have
calculated $\lan E\ran(\beta_{conf})$. Fig. \ref{edwards}
outlines a very good agreement between $\lan E \ran(\beta_{conf})$ and
${\overline E}(\beta_{fd})$ (notice that there are no adjustable
parameters).

Note that  the maximum energy reached by the system under the tap dynamics,
$E_{max}(\tau_0)\equiv {\overline E}(T_\Gamma\rightarrow\infty ,\tau_0)$,
is less than the maximum energy of the inherent states, $\lan E \ran(T_{conf}
\rightarrow\infty)$, for every value of $\tau_0$ considered.  Such a
prediction, which may have important practical consequences (e.g., in
powders technologies), is consistent with some experimental observations on
tapped granular materials \cite{Knight}, where the system density was shown to
approach asymptotically
a plateau value apparently higher than the minimal possible
packing density (obtained, for instance, by just pouring grains in the
container) even for very large tap amplitudes.

\begin{figure}
\mbox{\epsfysize=7cm\epsfbox{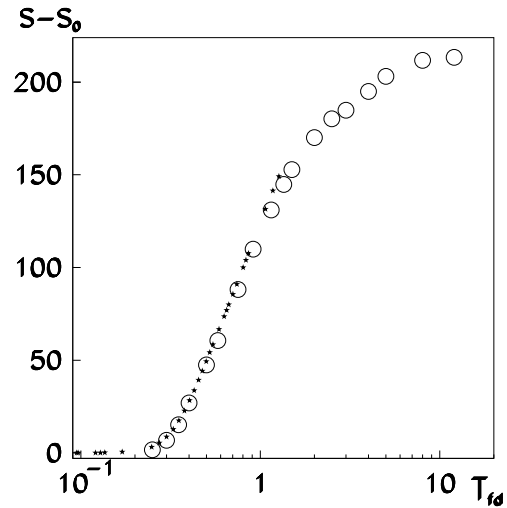}}
\caption{\label{entropia2}
The configurational entropy, $S-S_0$, (the
black empty circles in figure) as function of $T_{conf}$
(in units of $J$), compared with $S'(E)-S'_0\equiv \int_0^E
\beta_{fd}(E') dE'$ plotted as function of $T_{fd}$ (in units of $J$), in
the frustrated lattice gas model for $\rho=0.65$. The unknown non
negative constant $S_0$ is the entropy at $T_{conf}=0$.}
\end{figure}

Using eq.(\ref{omega}), we have finally evaluated the configurational entropy
as $S(E)-S_0=\int_0^E \beta_{conf}(E') dE'$  (where the unknown non negative
constant, $S_0\equiv S(E=0)$, is the entropy at $T_{conf}=0$).
In Fig. \ref{entropia2}, the configurational entropy, $S-S_0$, is plotted
as function of $T_{conf}$.
We have also evaluated the integral  $S'(E)-S'_0\equiv
\int_0^E \beta_{fd}(E') dE'$. In Fig. \ref{entropia2}, $S'-S'_0$  is plotted
as function of $T_{fd}$ and it is compared with the configurational entropy.
The agreement is again very good.

\subsection{Quasi-stationary case}
\label{outequ}
\begin{figure}
\mbox{\epsfysize=7cm\epsfbox{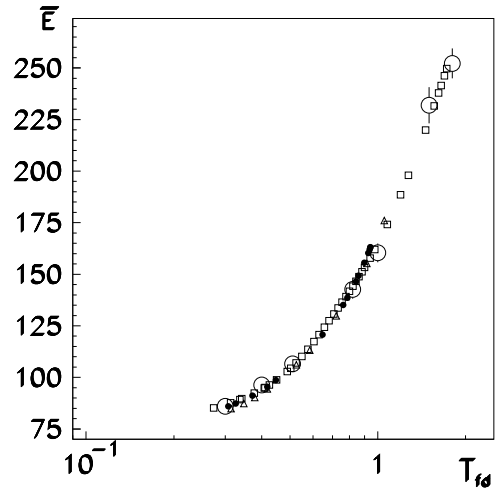}}
\caption{\label{edwards2}
The time average, ${\overline E}$, and the
ensemble average over the distribution eq.(\ref{pr}),
$\lan E \ran$ (the black empty circles), plotted respectively
as a function of $T_{fd}$  and $T_{conf}$ (in units of $J$), 
in the frustrated lattice
gas model, at $\rho=0.75$. As well as at $\rho=0.65$, 
there is a very good agreement between the two independently 
calculated sets of points. 
}
\end{figure}

We have also studied the frustrated lattice gas model for $\rho=0.75$.
Differently from the previous case, for small enough values of the tap 
amplitude $T_{\Gamma}$, the system does not reach a stationary state
during our observation time. In Fig. \ref{energia}, the energy, $E(t)$, of
the inherent states obtained for $T_\Gamma=0.3~J$ and $\tau_0=1~MCS$ is shown.
$E(t)$ now changes in time and 
the system is not in a stationary state; however, 
$E(t)$, at long times decays very slowly \cite{nota}.
In this regime the time averages are computed over a time
interval such that the energy is practically constant (in the case of Fig.
\ref{energia}, the time average is performed over the time interval 
$(3{\cdot} 10^5, 3{\cdot} 10^5+10^4)$).
Performing the same procedure described in the stationary case,
a collapse of data is again found (see Fig.\ref{edwards2}).

\begin{figure}
\mbox{\epsfysize=7cm\epsfbox{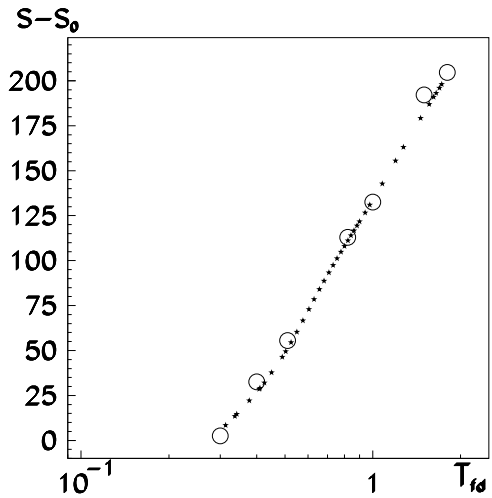}}
\caption{\label{entropia3}
The configurational entropy, $S-S_0$, (the
black empty circles in figure) as function of $T_{conf}$
(in units of $J$), compared with $S'(E)-S'_0\equiv \int_0^E
\beta_{fd}(E') dE'$ plotted as function of $T_{fd}$ (in units of $J$), in
the frustrated lattice gas model for $\rho=0.75$. The unknown non
negative constant $S_0$ is the entropy at $T_{conf}=0$.}
\end{figure}

We have again evaluated the configurational entropy, $S(E)-S_0=\int_{E_{min}}^E
\beta_{conf}(E') dE'$  (where the unknown non negative
constant, $S_0\equiv S(E_{min})$, is the entropy at $T_{conf}=0$
and $E_{min}$ is the minimum value of energy obtained \cite{nota4}).
In Fig. \ref{entropia3}, the configurational entropy, $S-S_0$, is plotted
as function of $T_{conf}$.
We have also evaluated the integral  $S'(E)-S'_0\equiv
\int_0^E \beta_{fd}(E') dE'$. In Fig. \ref{entropia3}, $S'-S'_0$  is plotted
as function of $T_{fd}$ and it is compared with the configurational entropy.
The agreement is again very good.

We cannot exclude that the agreement here found even for low energy may be due 
to the fact that the system, which is not in a stationary state, is however
very near stationarity.   

\section{A monodisperse hard-sphere system under gravity}
\label{hardsphere}

\begin{figure}
\includegraphics[scale=0.4,angle=-90]{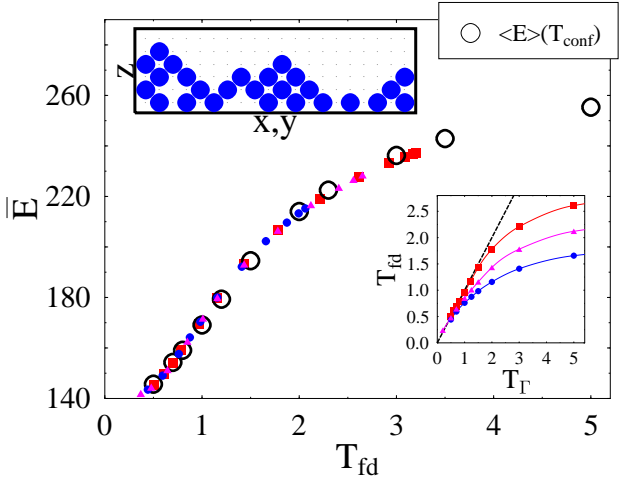}
\caption{\label{hard_sph}
{\bf Main frame}
The time average, ${\overline E}$, and the
ensemble average over the distribution eq.(\ref{pr}),
$\lan E \ran$ (the black empty circles), plotted respectively
as a function of $T_{fd}$  and $T_{conf}$ (in units $mga_0$),
in the 3D monodisperse hard-sphere system under gravity described in the text
(and schematically depicted in the {\bf upper inset}).
Time averages over the tap dynamics and Edwards' ensemble averages then 
coincide. 
{\bf Lower Inset}
The temperature $T_{fd}\equiv \beta^{-1}_{fd}$ defined by
eq.(\ref{Sb}) as function of $T_\Gamma$ (in units $mga_0$)
for $\tau_0=500,10,5~MCS$ (from top to bottom).
The straight line is the function $T_{fd}=T_\Gamma$.
}
\end{figure}

As a model more appropriate for granular media, we have also 
studied a system of monodisperse hard-sphere (with diameter $a_0=1$) under 
gravity, where the
centers of mass of grains are constrained to move on the sites of a cubic
lattice (see upper inset in Fig.~\ref{hard_sph}). 
The Hamiltonian of the system is:
\begin{equation}
{\cal H}= {\cal H}_{hc}(\{n_i\})+ gm \sum_i n_i z_i,
\label{H}
\end{equation}
where the height of site $i$ is $z_i$, $g=1$ is the gravity acceleration, 
$m=1$ the grains mass, $n_i$ is the usual occupancy variable and 
${\cal H}_{hc}(\{n_i\})$ is the hard core term preventing 
the overlapping of nearest neighbors grains 
(the analogy with eq.(\ref{flg}) can be appreciated by writing down  
${\cal H}_{hc}$: it can be written as 
${\cal H}_{hc}(\{n_i\})=J\sum_{\lan ij\ran }n_in_j$, where 
the limit $J\rightarrow\infty$ is taken).

\begin{figure}
\includegraphics[scale=0.4,angle=-90]{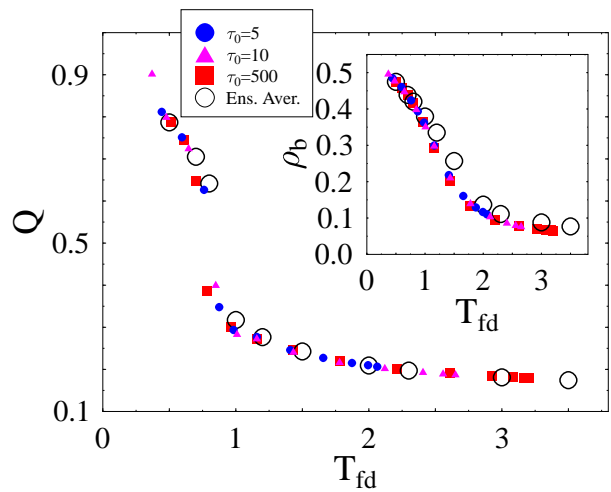}
\caption{\label{mimmo}
The density self-overlap function, $Q$, and ({\bf upper inset})
the system density on the bottom layer, $\rho_b$,
plotted as function of $T_{fd}$ (in units $mga_0$), compared with the
ensemble averages over the distribution eq.(\ref{pr})
(the black empty circles), plotted as function of $T_{conf}$ (in units
$mga_0$),
in the 3D monodisperse hard-sphere system under gravity. 
Also for this system there is a very good agreement between the 
independently calculated time averages over the tap dynamics and the 
statistical mechanics ensemble averages \'a la Edwards. 
}
\end{figure}

We have considered a system of $N=240$ particles, and 
performed a tap dynamics which allows the system to explore its
inherent states. We have considered $3$ different values
of the tap duration, $\tau_0=500,~10,~5~MCS$.
In this case we again obtain that the
asymptotic states reached by the system can be described by a single
thermodynamic
parameter, $\beta_{fd}$, evaluated by integration of eq. (\ref{Sb}).
We have moreover calculated the system density on the bottom layer, $\rho_b$,
and the density self-overlap function, $Q$, 
and verified that, when plotted as function
of $\beta_{fd}$, they scale on a single master function (see Fig. \ref{mimmo}).

As  described in Sect. \ref{EDW}, we have calculated the Edwards' averages as
function of $\beta_{conf}$. As we can see in Fig.\ref{hard_sph}, we obtain a
very good agreement between $\lan E \ran(\beta_{conf})$ and
${\overline E}(\beta_{fd})$. The same agreement is found for the other quoted
observables, $\rho_b$ and $Q$ (see Fig. \ref{mimmo}).
\section{A hard-sphere binary mixture under gravity}
\label{twotemp}
Finally we consider a hard-sphere binary system made of two species 1 (small)
and 2 (large) with grain diameters $a_0$ and $\sqrt{2} a_0$, under gravity
on a cubic lattice of spacing $a_0=1$.
We set the units such that the two kinds of grain have masses
$m_1=1$ and $m_2=2$, and the gravity acceleration is $g=1$. The hard core
potential ${\cal H}_{hc}$ is such that two large nearest neighbor particles
cannot overlap. This implies that only couples of small particles can be
nearest neighbors on the lattice. The system overall Hamiltonian is:
\begin{equation}
{\cal H}={\cal H}_{hc}+ m_1gH_1 + m_2gH_2,
\label{HSM}
\end{equation}
where $H_1=\sum_{i}^{(1)}z_{i}$ and  $H_2=\sum_{i}^{(2)}z_{i}$, the
height of site $i$ is $z_i$ and the two sums are over all particles of
species 1 and 2 respectively. In the above units, the gravitational
energies in a given configuration are thus $E_1=H_1$ and $E_2=2H_2$.

Grains are confined in a box of linear size $L$ between hard walls and
periodic boundary conditions in the horizontal directions.
$N_1=120$ grains of type 1 and $N_2=40$ grains of type 2 are initially prepared
in a random loose stable pack. Under the tap dynamics the system
approaches a stationary state for each value of the tap parameters
$T_{\Gamma}$ and $\tau_0$.
\begin{figure}
\includegraphics[scale=0.4,angle=-90]{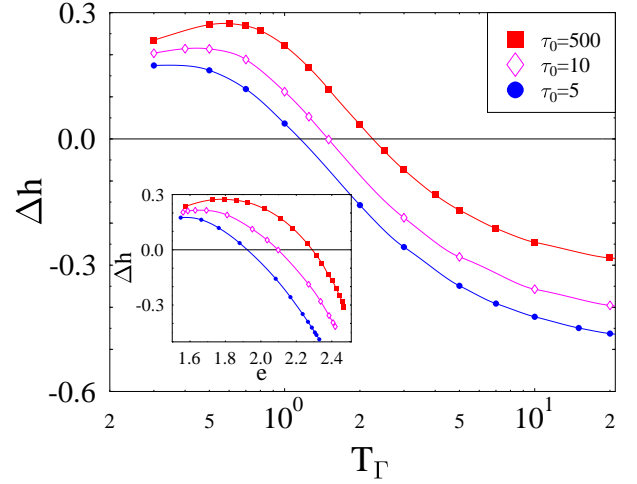}
\caption{{\bf Main frame} The difference of the average heights of small
and large grains, $\Delta h=h_1-h_2$, measured at stationarity 
in the binary hard spheres mixture under gravity, 
is plotted as a function of tap amplitude, $T_{\Gamma}$
(in units $mga_0$). The three sets of
points correspond to the shown tap durations, $\protect\tau_0$.
At high $T_{\Gamma}$ larger grains are found above the smaller, i.e,
$\Delta h<0$, as in the Brazil nut effect (BNE).
Below a  $T^*_{\Gamma}(\tau_0)$ the opposite is found
(Reverse Brazil nut effect, RBNE).
{\bf Inset}
The $\Delta h$ data of the main frame are plotted as a function of the
corresponding average energy, $e$. The three sets of data do not collapse onto
a single master function, showing that a single macroscopic observable, such
as $e$, doesn't characterize the system status.
}
\label{fig_dh}
\end{figure}
\begin{figure}
\includegraphics[scale=0.4,angle=-90]{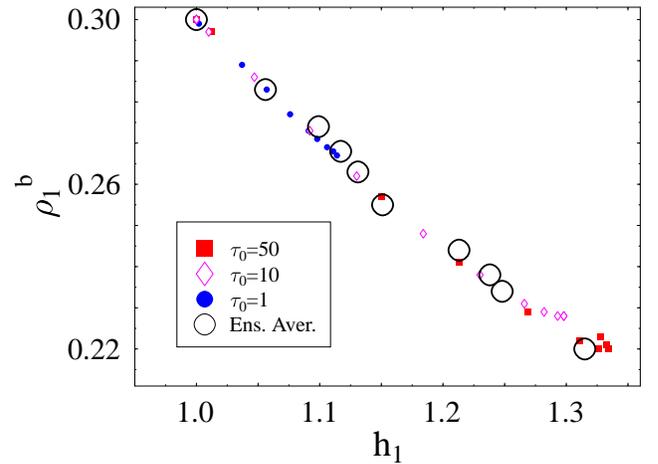}
\caption{The average density of small grains on the
box bottom layer, $\protect\rho_1^{b}$, measured at
stationarity as a function of the height of small particles, $h_1$, 
in the binary hard spheres mixture under gravity. Data
corresponding to different $T_{\Gamma}$ and $\protect\tau_0$ approximately
scale on a single master function.
The empty circles are the corresponding values obtained by ensemble
average with the two temperatures Gibbs measure proposed in the text.
}
\label{fig_qr2nc_1}
\end{figure} 
\begin{figure}
\includegraphics[scale=0.4,angle=-90]{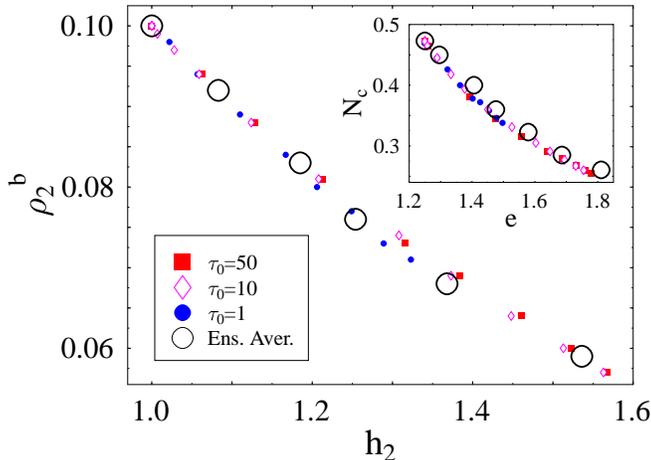}
\caption{
{\bf Main frame} The average density of
large grains on the box bottom layer, $\protect\rho_2^{b}$, obtained for
different $T_{\Gamma}$ and $\protect\tau_0$, scale almost on a single master
function when plotted as a function of the large grains height, $h_2$.
{\bf Upper inset} The average number of contacts between large grains
per particle, $N_c$, obtained for different $T_{\Gamma}$ and $\protect\tau_0$,
scale on a single master function when plotted as a function of the system
energy, $e$. }
\label{fig_qr2nc_2}
\end{figure}
In Fig.\ref{fig_dh}, we plot as function of $ T_{\Gamma}$ (for several values
of $\tau_0$) the asymptotic value of the
{\em vertical} segregation parameter, i.e.,
the difference of the average heights of the small and large grains at
stationarity, $\Delta h(T_{\Gamma},\tau_0)\equiv h_1-h_2$.
Here $h_1$ and $h_2$ are the average of $H_1/N_1$ and $H_2/N_2$
over the tap dynamics in the stationary state.
(An interpretation, in terms of the approach here presented, 
of the size segregation phenomenon here found and experimentally 
observed in a hard-spheres binary mixture under gravity is given 
in Ref. \cite{NFC}). 

The results given in the main panel of Fig.~\ref{fig_dh}
apparently show that $T_{\Gamma}$ is not a
right thermodynamic parameter, since sequences of taps with
different $\tau_0$ give different values for the system observables.
However, if the stationary states corresponding to different tap dynamics
(i.e., different $T_{\Gamma}$ and $\tau_0$), are indeed characterized by a
single thermodynamic parameter, the curves of
Fig.\ref{fig_dh} should collapse onto a universal master function when
$\Delta h(T_{\Gamma},\tau_0)$ is parametrically plotted as function of an
other macroscopic observable such as the average energy,
$e(T_{\Gamma},\tau_0)=(E_1 + E_2)/N$ ($N$ is the total number of particles).
This collapse of data is
clearly not observed here, as it is apparent in the inset of
Fig.\ref{fig_dh}.
We show, instead, that two macroscopic quantities may be sufficient to
characterize uniquely the stationary state of the system. These two
quantities are, for instance,
the energy $e$ and the height difference $\Delta h$. Of
course since $e = ah_1 + 2bh_2$ (where $a= N_1/N$ and $b=N_2/N$) and
$\Delta h = h_1-h_2$, we can also choose $h_1$ and $h_2$ to characterize the
stationary state.
Namely, we show that any macroscopic
quantity $A$, averaged over the tap dynamics in
the stationary state, is only dependent on $h_1$ and $h_2$,
i.e., $A= A(h_1,h_2)$. We have checked that this is the case for
several independent observables, such as
the number of contacts between large particles, $N_{c}$, the density
of  small and large particles on the bottom layer, $\rho_{1}^{b}$ and
$\rho_{2}^{b}$, and others.
In particular,
as shown in Fig.s \ref{fig_qr2nc_1} and \ref{fig_qr2nc_2}, we find
with good approximation that: $N_{c}\simeq N_{c}(e)=N_{c}(ah_1+bh_2)$,
$\rho_{2}^{b}\simeq\rho_{2}^{b}(h_{2})$, $
\rho _{1}^{b}\simeq \rho _{1}^{b}(h_{1})$.
Therefore we need both $h_{1}$ and $h_{2}$ 
to characterize unambiguously the state of the system;
namely all the observables assume the same values in a stationary
state characterized by the same values of $ h_{1}$ and $h_{2}$,
independently on the previous history (i.e., in our case independently on the
particular tapping parameters $T_{\Gamma}$ and $\tau_0$).

We again find that the stationary state can be genuinely considered as a
thermodynamic state. Therefore we can ask what is the probability
distribution, $P_r$, of finding the system in the inherent state $r$
corresponding to an
energy $E_{1r}$ for the small particles and $E_{2r}$ for the large
particles. We again assume that the microscopic distribution is given by the
principle of maximum entropy $S=-\sum_r P_r\ln P_r$, now under the
condition
that the average energy $E_1 =\sum_r P_r E_{1r} $ and
$E_2 =\sum_r P_r E_{2r} $ are independently fixed.
This can be done by introducing two Lagrange multipliers
$\beta_1$ and $\beta_2$, which are  determined by the constraint on $ E_1$
and $E_2$, and can be respectively considered as the ``inverse configurational
temperature'' of species 1 and 2. This procedure leads to the Gibbs
result:
\begin{eqnarray}
P_r=\frac{e^{-\beta_1 E_{1r} - \beta_2 E_{2r}}}{Z}
\label{pr2}
\end{eqnarray}
where
$Z=\sum_r \exp\left[-\beta_1 E_{1r} - \beta_2 E_{2r}\right]$
and, in the thermodynamic limit, the entropy, $S$, is given by:
\begin{eqnarray}
S = \ln \Omega (E_1,E_2),
\label{omega2}
\end{eqnarray}
and $\beta_1$ and $\beta_2$:
\begin{eqnarray}
\beta_1 = \frac{\partial \ln \Omega (E_1,E_2) }{\partial E_1},\quad
\beta_2 = \frac{\partial \ln \Omega (E_1,E_2) }{\partial E_2}.
\label{beta}
\end{eqnarray}
Here $\Omega (E_1,E_2)$ is the number of inherent states corresponding to
energy $E_1$ and $E_2$.
The hypothesis that the ensemble distribution at stationarity is given
by eq.(\ref{pr2}) can be tested as follows. We have to check that
the time average of any quantity,
$A(h_{1},h_{2})$, as
recorded during the taps sequences in a stationary state characterized
by given values  $h_{1}$ and $h_{2}$, must coincide with the ensemble
average, $\langle A\rangle (h_{1},h_{2})$, over the distribution
eq.(\ref{pr2}). To this aim, we have calculated the
ensemble averages $\langle N_{c}\rangle $,  $\langle \rho_{2}^{b}\rangle$,
$\langle \rho _{1}^{b}\rangle $ for different values of $\beta_1$ and
$\beta_2$;we have expressed parametrically  $\langle N_{c}\rangle $, $\langle
\rho _{2}^{b}\rangle $, $\langle \rho _{1}^{b}\rangle $,
as function of the average of $ h_{1}$ and $ h_{2}$,
and compared them with the corresponding quantities, $N_{c}$, $\rho_{1}^{b}$
and $\rho _{2}^{b}$, averaged over the tap dynamics.
The two sets of data are plotted in 
Fig.s \ref{fig_qr2nc_1} and \ref{fig_qr2nc_2} showing a
good agreement (notice, there are no adjustable parameters).
In order to calculate the ensemble averages we simulate
the model with $\cal H$ from eq.(\ref{HSM})
where we impose the constraint that the only accessible states are the
inherent states, as already described in Sect. \ref{EDW}.

\section{Conclusions}

In conclusion, in the context of models for glasses and granular materials we
have shown that the stationary (or quasi-stationary) state reached by the
system subject to a tap dynamics among its inherent states is genuinely 
a thermodynamic state, which can be well described by Edwards' assumption of 
an uniform measure, i.e., a probability distribution 
obtained assuming the validity of the principle of maximum entropy.
In particular in the frustrated lattice gas model and in the system of 
monodisperse hard-spheres under gravity we have found that
the observables recorded during different tap
sequences (different amplitude and duration of taps) fall on universal master
curves when plotted as a function of a single thermodynamic parameter.
These curves turn out to coincide 
with those predicted, within the described Statistical Mechanics approach, 
by the generalized Gibbs distribution of 
eq.(\ref{pr}). On the other hand, the results obtained in
a system under gravity made of particles of two different sizes
show that a single thermodynamic parameter is
not enough to describe the macrostates, and two
configurational temperatures are instead necessary.
In general, for a more complex system one might expect more constraints to be
imposed, leading to more than two thermodynamic parameters
\cite{berg,lefevre,NFC}.
In practice, the criteria to determine a priori the required parameters
can be not easily accessible. However more recently we have extended
data regarding the hard-sphere binary mixture for
very low energy \cite{CFN} and found that only one thermodynamical parameter is
necessary to describe the stationary state. This seems to be a general feature
\cite{NCF1}. If this is the case, a Statistical Mechanics approach  with only
one thermodynamical variable may be feasible for low energy.

\begin{acknowledgments}
This work was partially supported by the TMR-ERBFMRXCT980183, INFM-PRA(HOP),
MURST-PRIN 2000 and MIUR-FIRB 2002. The allocation of computer resources from
INFM Progetto Calcolo Parallelo is acknowledged.
\end{acknowledgments}

\appendix
\section{Determination of the integration constant $\beta_0$} 
Adapting to lattice models the procedure of Sciortino {\em et al.}
\cite{sciortino}, we
have evaluated $\beta_{fd}$ at small values of $T_\Gamma$,
for $\tau_0\rightarrow\infty$, and consequently the integration constant,
$\beta_0$.

\begin{figure}
\mbox{\epsfysize=7cm\epsfbox{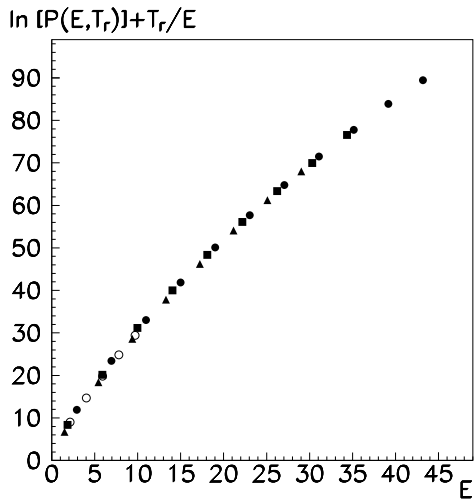}}
\caption{\label{entropia1}
The curves $\ln \left  [P(E,T_\Gamma)\right ] +E/T_\Gamma$ (apart from a
$T_\Gamma-$dependent constant) as
function of the energy $E$ in the frustrated lattice gas model for $\rho=0.65$
and $T_\Gamma=0.275,~0.425,~0.475,~0.525~J$.}
\end{figure}

Given an inherent state, $r$, of energy, $E_r$, we define the basin of
attraction, $B_r$, of such state $r$  as the set of states in the
configurational space, which after the quench at $T=0$ are frozen in the
inherent state $r$. Therefore the  probability distribution, $P_r$, of
finding the system in the inherent state $r$ after quenching the system from an
equilibrium state at temperature $T_\Gamma$, can be written as
\begin{equation}
P_r=\frac{\sum_{r'} e^{-E_{rr'}/T_\Gamma}}{Z_{G}(T_\Gamma)},
\label{basin1}
\end{equation}
where $Z_{G}(T_\Gamma)$ is the partition function of the system in
equilibrium at temperature $T_\Gamma$ and $\sum_{r'}$ is the the sum over
all the states $r'$ belonging to the basin $B_r$ of energy $E_{rr'}$. By
putting $E_{rr'}=E_r+\Delta_{rr'}$, the distribution (\ref{basin1}) can be
written as
\begin{equation}
P_r=\frac{e^{-(E_{r} + g_r(T_\Gamma))/T_\Gamma}}{Z_{G}(T_\Gamma)},
\label{basin2}
\end{equation}
where $e^{- g_r(T_\Gamma)/T_\Gamma} = \sum_{r'} e^{-\Delta_{rr'}/T_\Gamma}$. 
From eq. (\ref{basin2}) it follows that the probability of finding the system
in any inherent state of energy $E$, $P(E,T_\Gamma)=\sum_r P_r$ (where
$\sum_r$ is the sum over all the inherent states $r$ of energy $E$) is
given by
\begin{equation}
P(E,T_\Gamma)=\frac{\Omega(E) e^{-E/T_\Gamma}
e^{-f(T_\Gamma,E)/T_\Gamma}}{Z_{G}(T_\Gamma)},
\label{lab1}
\end {equation}
where $\Omega(E)$ is the number of inherent states of energy $E$ and $
e^{-f(T_\Gamma,E)/T_\Gamma}=\frac{\sum_r e^{-
g_r(T_\Gamma)}/T_\Gamma}{\Omega(E)}$.  From eq. (\ref{lab1}):
\begin{equation}
\ln \left  [P(E,T_\Gamma)\right ]+\frac{E}{T_\Gamma}= -\frac{f(T_\Gamma,E)}{T_\Gamma}+
\ln \left[\Omega(E)\right ]-\ln \left[Z_{G}(T_\Gamma)\right ].
\label{lab2}
\end {equation}
The probability distribution of finding the system in any inherent state
of energy $E$, $P(E,T_\Gamma)$, is measured during the tap dynamics with
amplitude $\tau_0\rightarrow \infty$.  If $f(T_\Gamma,E)$ has only a weak 
dependence on $E$, then it is possible to superimposed the curves, $\ln \left
[P(E,T_\Gamma)\right ] +E/T_\Gamma$, at different $T_\Gamma$ which overlap in
$E$ by adding a $T_\Gamma-$dependent constant. 
This result is obtained for $T_\Gamma\le 0.525$, as shown in
Fig.\ref{entropia1}, and suggests that in this interval $f(T_\Gamma,E) \simeq
f(T_\Gamma)$. If this is the case, from eq.(\ref{lab1}) it follows:  
\begin{equation}
P(E,T_\Gamma)\simeq \frac{\Omega(E) e^{-E/T_\Gamma}}{Z(T_\Gamma)},
\label{lab3}
\end {equation}
where
\begin{equation}
Z(T_\Gamma) = e^{f(T_\Gamma)/T_\Gamma}Z_{G}(T_\Gamma) =
\sum_E\Omega(E) e^{-E/T_\Gamma}.
\label{lab4}
\end {equation}
The last equality stems from the normalization condition on
$P(E,T_\Gamma)$.

\begin{figure}
\mbox{\epsfysize=7cm\epsfbox{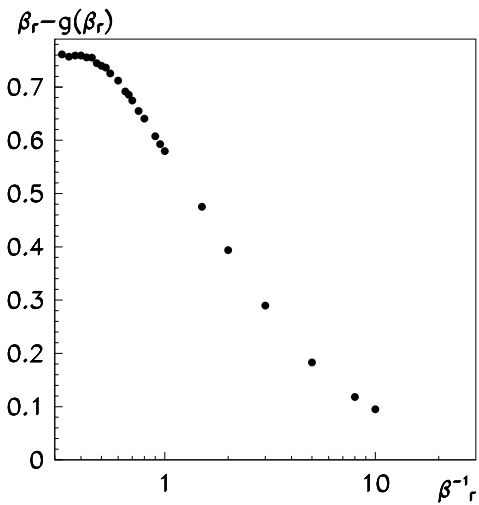}}
\caption{\label{beta0}
The curve $\beta_\Gamma-g(\beta_\Gamma)=\beta_\Gamma-\left
[\beta_{fd}(\beta_\Gamma)-\beta_0\right ]$
as function of $\beta^{-1}_\Gamma$ (in units of $J$) 
in the frustrated lattice gas model for
$\rho=0.65$. The limit $\beta^{-1}_\Gamma\rightarrow 0$ of
$\beta_\Gamma-g(\beta_\Gamma)$
gives the integration constant $\beta_0$.}
\end{figure}

From eq.(\ref{lab3}) it follows that at small $T_\Gamma$,
$\beta_\Gamma  \equiv T^{-1}_\Gamma$ satisfies eq.(\ref{Sb}). Therefore at
small $T_\Gamma$, $\beta_{fd}$ and $\beta_\Gamma$ coincide. 
The constant $\beta_0$ is consequently obtained as the limit,
for $T_\Gamma\rightarrow 0$, of the function $\beta_\Gamma -
g(\beta_\Gamma)$ (see Fig.\ref{beta0}).

\begin{thebibliography}{999}
\bibitem{Stillinger}
F.H. Stillinger and T.A. Weber, Phys. Rev. A {\bf 25}, 978 (1982);
Science {\bf 225},  983 (1984).
F. H. Stillinger,  Science, {\bf 267} 1935, (1995).
S. Sastry, P.G. Debenedetti, F.H. Stillinger, Nature {\bf 393}, 554 (1998).
\bibitem{parisi}
B. Coluzzi, G. Parisi and P. Verrocchio, Phys. Rev. Lett. {\bf 84}, 306 (2000).
\bibitem{sciortino}
F. Sciortino, W. Kob, P. Tartaglia, Phys. Rev. Lett. {\bf 83}, 3214 (1999).

\bibitem{tartaglia}
W. Kob, F. Sciortino and P. Tartaglia, Europhys. Lett., {\bf 49}, 590 (2000).\\
F. Sciortino and P. Tartaglia, Phys. Rev. Lett. {\bf 86}, 107 (2001).

\bibitem{JNBHM}
  H.M. Jaeger, S.R. Nagel and R.P. Behringer,
  Rev. Mod. Phys. {\bf 68}, 1259 (1996).

\bibitem{nota_frozen} Grains are
``frozen'' because, due to their large masses and dissipation 
\cite{JNBHM}, the thermal
kinetic energy is negligible compared to the gravitational energy; thus
the external bath temperature, $T_{bath}$, can be considered equal to
zero.

\bibitem{Edwards}
S.F. Edwards and R.B.S. Oakeshott, Physica A {\bf 157}, 1080 (1989).
A. Mehta and S.F. Edwards, Physica A {\bf 157}, 1091 (1989).
S.F. Edwards, in {\em ``Disorder in Condensed Matter Physics''} page. 148,
Oxford Science Pubs (1991); and
in {\em Granular Matter: an interdisciplinary approach},
(Springer-Verlag, New York, 1994), A. Mehta ed.
\bibitem{peliti} L. F. Cugliandolo, J. Kurchan and L. Peliti,
Phys. Rev. E {\bf 55},3898 (1997). 

\bibitem{franzvirasoro} 
R. Monasson, Phys. Rev. Lett. {\bf 75}, 2847 (1995).
Th.M. Nieuwenhuizen, Phys. Rev. E {\bf 61}, 267 (2000).
S. Franz and M.A. Virasoro, J. Phys. A {\bf 33}, 891 (2000).
A. Crisanti and F. Ritort, 
Journal of Chemical Physics {\bf 113}, 10615 (2000).
J. Kurchan, {\em cond-mat/9812347}; and
in ``Jamming and Rheology: Constrained Dynamics on
Microscopic and Macroscopic Scales'', A.J. Liu and S.R. Nagel Eds.,
Taylor and Francis,  London (2001).  

\bibitem{n1} M. Nicodemi, Phys. Rev. Lett. {\bf 82}, 3734 (1999).

\bibitem{barrat} A. Barrat, J. Kurchan, V. Loreto and
M. Sellitto, Phys. Rev. Lett. {\bf 85}, 5034 (2000); Phys. Rev. E {\bf
63},051301 (2001)
\bibitem{Makse} H. A. Makse and J. Kurchan, Nature {\bf 415}, 614 (2002).  

\bibitem{NC} A. Coniglio and M. Nicodemi, Physica A {\bf 296}, 451 (2001).
 
\bibitem{NCF1} A. Coniglio, A. Fierro and M. Nicodemi, Physica A {\bf 302}
(1-4): 193 (2001);  A. Fierro, M. Nicodemi and A. Coniglio, 
Europhys. Lett. {\bf 59}, 642 (2002); A. Coniglio, A. Fierro and M. Nicodemi,
Europ. Phys. Jour. E, in press (2002). 

\bibitem{NFC} M. Nicodemi, A. Fierro and A. Coniglio, {\em cond-mat}/0202500
and Europhys. Lett. in press.

\bibitem{brey}
J.J. Brey, A. Prados, B. S\'{a}nchez-Rey, Physica A {\bf 275}, 310 (2000);
A. Prados, J.J. Brey, B. S\'{a}nchez-Rey, Physica A {\bf 284}, 277 (2000).

\bibitem{Dean} D. S. Dean and A. Lef\`{e}vre,
Phys. Rev. Lett. {\bf 86}, 5639 (2001); J. Phys. A: Math. Gen. {\bf 34},
L213 (2001); {\em cond-mat}/0106220.
\bibitem{berg} J. Berg, S. Franz and M. Sellitto, 
Eur. Phys. J. B {\bf 26}, 349 (2002).
\bibitem{NCH} M. Nicodemi, A. Coniglio, H.J. Herrmann,
Phys. Rev. E {\bf 55}, 3962 (1997); J. Phys. A {\bf 30}, L379 (1997).
A. Coniglio and H.J. Herrmann, Physica A {\bf 225}, 1 (1996).
\bibitem{mehta} J. Berg, A. Mehta, Europhys. Lett. {\bf 56}, 784 (2001).

\bibitem{cugliandolo} L. Berthier, L.F. Cugliandolo and J.L. Iguain, 
Phys. Rev. E {\bf 63}, 051302 (2001).  

\bibitem{nota_2t} the possibility of introducing more than one thermodynamic
parameter has been also  suggested in \cite{berg} and recently discussed in
the context of a Constrained Ising Chain in \cite{lefevre}.

\bibitem{lefevre} A. Lef\`{e}vre, {\em cond-mat}/0202376. 


\bibitem{barkermehta}
A. Mehta and G. C. Barker, Phys. Rev. Lett. {\bf 67}, 394 (1991).

\bibitem{dfnc}
A. Coniglio, A. de Candia, A. Fierro, M. Nicodemi,
Jour. Phys.: Cond. Mat. {\bf 11}, A167 (1999).
\bibitem{varenna} A. Coniglio, ``Frustration and connectivity in glass forming
systems and granular materials", in Proc. Int. School on the Physics of Complex
Systems ``Enrico Fermi" CXXXIV course (Varenna 1996), edited by F. Mallamace and
H.E.  Stanley (IOS Press, Amsterdam,1997), p.491.
M. Nicodemi and A. Coniglio, J. Phys A Lett. {\bf 30}, L187 (1996).
F. Ricci-Tersenghi, D.A. Stariolo and J.J. Arenzon, Phys. Rev. Lett. {\bf 84},
4473 (2000).
\bibitem{our_rev} A. Coniglio, M. Nicodemi, J. Phys.: Cond. Matt. {\bf 12}, 6601
(2000); M. Nicodemi, A. Coniglio, Phys. Rev. Lett. {\bf 82}, 916 (1999).
\bibitem{jef} J.J. Arenzon, J. Phys. A. {\bf 32}, L107 (1999).
\bibitem{antonio} A. de Candia and A. Coniglio, Phys. Rev. E {\bf 65}, 16132
(2002).
\bibitem{nota1} $\tau_0$ is measured in Monte Carlo steps (MCS), where $1~MCS$
corresponds to $N$ attempts to move a particle randomly chosen, and $N=\sum_i
n_i$ is the number of particles.
\bibitem{nota2}
Note that with our approach it is also possible to explore low density inherent
states in a stationary regime and not only the off-equilibrium ``glassy
regime'', as instead in Ref. \cite{barrat}. For instance, the frustrated
lattice gas model at density $\rho=0.65$ is hardly found in an out of
equilibrium quasi-stationary state (at any finite value of the bath temperature
the system
quickly reaches the equilibrium state).
We have considered the model eq.(\ref{flg}), at $\rho=0.65$, and we have
performed an usual Monte Carlo diffusive dynamics of the model. At 
a starting time we prepared 
the system in equilibrium with a very high bath temperature. 
Afterwards, it is suddenly cooled at a very low $T_{bath}$, but  
the system quickly
reaches the equilibrium state, and $T_{dyn}$ coincides with $T_{bath}$. 
\bibitem{nota3}       
Where not explicitely shown, the error bars of the data are more or less equal 
to the size of symbols in figures.
\bibitem{Knight} J.B. Knight, 
C.G. Fandrich, C.N. Lau, H.M. Jaeger and S.R. Nagel,
Phys. Rev. E {\bf 51}, 3957 (1995).
E.R. Nowak, J.B. Knight, E. Ben-Naim, H.M. Jaeger and S.R. Nagel,
Phys. Rev. E {\bf 57}, 1971 (1998).
E.R. Nowak, J.B. Knight, M. Povinelli, H.M. Jaeger, and S.R. Nagel,
Powder Technology {\bf 94}, 79 (1997).                                         

\bibitem{nota} In both cases, $\rho=0.65$ and $0.75$,
the stationary self scattering functions obtained
for  finite tap duration $\tau_0$, are well fitted by stretched exponentials,
$e^{-(t/\tau)^a}$. The characteristic time scale, $\tau$, obtained in this way
increases as $T_\Gamma$ decreases, and seems to diverge only in the limit
$T_\Gamma\rightarrow 0$.

\bibitem{nota4} In the previous case, $\rho=0.65$, the minimum energy,
$E_{min}$, is zero.

\bibitem{CFN} A. Coniglio, A. Fierro, and M. Nicodemi (unpublished). 

\end{thebibliography}
\end{document}